\documentclass[journal=jacsat,manuscript=article]{achemso}

\usepackage[version=3]{mhchem} % Formula subscripts using \ce{}
\usepackage{graphicx}% Include figure files
\usepackage{dcolumn}% Align table columns on decimal point
\usepackage{bm}% bold math
\usepackage{braket}
\usepackage{enumitem}
\usepackage{color}
\usepackage{booktabs} 
\usepackage{newfloat}
\usepackage{algcompatible}
\usepackage[size=small]{caption}
\usepackage{etoolbox}
\usepackage{hyperref}

\author{Denis G. Artiukhin}
\affiliation[Freie Universität Berlin]
{Institut für Chemie und Biochemie, Freie Universität Berlin, \\ Arnimallee 22, 14195 Berlin, Germany.}
\email{denis.artiukhin@fu-berlin.de}

\title[NEO-sDFT]
  {Nuclear-Electronic Orbital \\ Subsystem Density Functional Theory}

\abbreviations{}

\begin{document}

\begin{tocentry}
\centering
\includegraphics[width=0.9\textwidth]{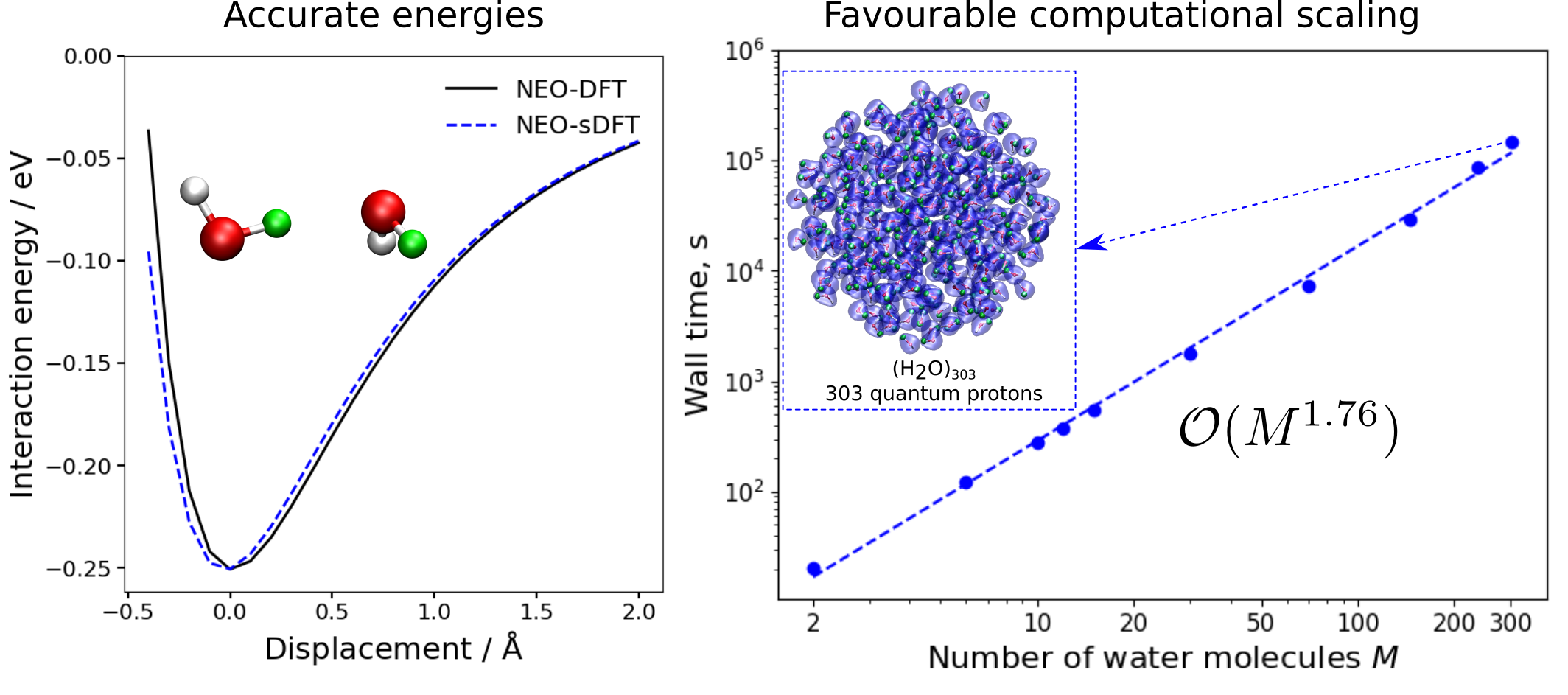}
\end{tocentry}

\begin{abstract}
We present a new computational approach based on Nuclear-Electronic Orbital and Subsystem Density Functional Theory. The resulting NEO-sDFT methodology enables quantum chemical calculations of large molecular systems composed of thousands of atoms while also describing selected protons quantum mechanically. A key feature of the associated program implementation is its flexibility in assigning quantum protons to subsystems, allowing different embedding schemes to be employed within a unified framework. The accuracy of NEO-sDFT is assessed by computing energies of water clusters of growing size. Deviations from reference Nuclear-Electronic Orbital Density Functional Theory results amount to only a few meV for water dimers near the equilibrium separation and are comparable to those of conventional Subsystem Density Functional Theory for larger clusters.
The favorable computational scaling of NEO-sDFT is further showcased through calculations on water clusters of increasing size, with up to 303 protons being treated quantum mechanically. Our results demonstrate that NEO-sDFT is an accurate and efficient approach for incorporating nuclear quantum effects into large-scale simulations, providing a foundation for future applications to proton transfer processes in biochemical systems.
\end{abstract}

\newpage
\clearpage

\section{Introduction} \label{sec:intro}

Nuclear-electronic orbital (NEO) approaches~\cite{neo_review2020,neo_review2021} constitute a family of multicomponent quantum chemical methods~\cite{Thomas1969,Tachikawa1998,Tachikawa1998_2,Shigeta1998,Gidopoulos1998,Kreibich2001} in which a few selected nuclei, typically protons, are treated quantum mechanically and on the very same footing as electrons. In this way, nuclear quantum effects such as zero-point energy are directly incorporated into the resulting energy value computed in a single point calculation. Hence, NEO provides an attractive route for simulations of molecular systems, where nuclear quantum effects play an important role, while avoiding extensive single-point energy evaluations on potential energy surfaces. One possible application of NEO methods is for the description of proton transfer processes (for examples, see Refs.~\citenum{Hazra2009,Yu2020}). However, the direction and rate of these reactions can be strongly affected by the interaction with surrounding molecules. Therefore, realistic simulations require the consideration of large environments, which presents a significant challenge, as the cost of quantum chemical approaches increases rapidly with the size of the molecular system considered.

Many previous studies have aimed at overcoming the aforementioned cost-associated limitation of NEO-based approaches. 
Early applications of Nuclear-Electronic Orbital Density Functional Theory (NEO-DFT) relied on electron--proton correlation functionals derived from the electron--proton pair density of explicitly correlated nuclear--electronic wave functions~\cite{chakraborty2008,chakraborty2009,sirjoosingh2011,sirjoosingh2012}, whose computations were very expensive. Subsequently, in close analogy to the local-density approximation (LDA) and generalized gradient approximation (GGA) families of conventional Density Functional Theory (DFT), electron--proton correlation functionals depending on the electron and proton densities were proposed~\cite{yang2017,brorsen2017,brorsen2018,tao2019}, constituting a major step towards cost-efficient NEO-DFT while still enabling accurate calculations of proton densities. These functionals were later extended to hybrid~\cite{Holzer2024} and double-hybrid~\cite{Hasecke2025} forms and provided the required foundation for further developments of NEO-DFT-type methods.
A combination of Quantum Mechanical NEO approaches with Molecular Mechanics (NEO-QM/MM)~\cite{chow2023} made it possible to describe a small region of interest with NEO methods while accounting for the effects of a large environment treated at the MM level, whereas the extension of NEO towards periodic computations~\cite{xu2022}
enabled the treatment of condensed-phase systems. It is also worth mentioning that considerable effort has been devoted to the development of more efficient computational algorithms. Examples include, but are not limited to, local-density fitting techniques~\cite{Hasecke2023,Hasecke2_2025}, self-consistent-field (SCF) initial guesses for quantum protons~\cite{Aodong2022,Feldmann2023,Hasecke2023,Lehtola2025,artiukhin2026}, and various SCF optimization strategies~\cite{Aodong2022,Feldmann2023,Lehtola2025,Greiner2026}.

Another approach, which can be employed for reducing the computational scaling of NEO-DFT, as was previously pointed out in Ref.~\citenum{artiukhin2024}, is Subsystem Density Functional Theory (sDFT)~\cite{senatore1986,johnson1987,cortona1991,jacob2014,jacob2024}. Conventional electronic sDFT partitions the total molecular system into subsystems based on the electron density. Each subsystem is then treated independently, while the interactions between subsystems are accounted for through the so-called embedding potential. This considerably reduces the computational cost compared to the supersystem DFT computation and allows one to calculate molecular systems composed of several thousand atoms~\cite{konig2013,arti2020a}.
In the context of NEO methods, the (NEO-DFT)-in-DFT embedding methodology has already been presented in Ref.~\citenum{culpitt2016}. 
However, this contribution predated the development of accurate electron--proton correlation functionals and was computationally very demanding due to the need to consider electron--proton pair 
densities associated with explicitly correlated wavefunctions.
Consequently, only applications to triatomic molecules were presented~\cite{culpitt2016}. Furthermore, as discussed later in this work, several methodological choices made while deriving this theory resulted in unfavorable computational scaling further limiting its applicability. Finally, to the best of our knowledge, an efficient and publicly available program implementation of this approach has not been reported to date.

In this work, we address the aforementioned limitations by deriving a new Nuclear-Electronic Orbital Subsystem Density Functional Theory (NEO-sDFT) for cost-efficient computations of large molecular systems and present its program code implementation. A central feature of the current formulation is its high flexibility in how quantum protons can be assigned to the active and environment subsystems. More specifically, either subsystem can be described using DFT or NEO-DFT, allowing for all combinations of ($X$)-in-($Y$), where $X$, $Y=$ DFT and NEO-DFT, embedding schemes within a unified framework. We demonstrate the robustness and accuracy of this approach by performing a number of proof-of-principle calculations on molecular systems ranging from small water dimers to large water clusters containing hundreds of molecules.

This work is organized as follows. First, theoretical foundations of NEO-DFT and derivations of new NEO-sDFT are presented. Subsequently, the computational protocol is described, followed by proof-of-principle computations and assessment of the resulting approach. The work concludes with a summary of main findings.

\section{Theory} \label{sec:theory}

In what follows, we first briefly outline the route to treat protons quantum mechanically within the DFT framework in Section~\nameref{sec:neodft} as was proposed in Refs.~\citenum{pak2007,chakraborty2008,chakraborty2009,sirjoosingh2011,sirjoosingh2012}. Subsequently in Section~\nameref{sec:neosdft}, we formulate NEO-sDFT
and compare it against previously proposed NEO-embedding theory from Ref.~\citenum{culpitt2016}.

\subsection{NEO-DFT} \label{sec:neodft}

The underlying idea of NEO approaches~\cite{webb2002} is to treat a few selected protons as quantum particles and on the very same footing as electrons. In the context of DFT, this results in the energy expression of the form~\cite{chakraborty2008},
\begin{multline} \label{eq:neo_energy}
E [ \rho^{\mathrm{e}}, \rho^{\mathrm{p}}  ] = T_{\mathrm{s}}^{\mathrm{e}} [ \rho^{\mathrm{e}}] + V_{\mathrm{nuc}}^{\mathrm{e}} [\rho^{\mathrm{e}}] +
J^{\mathrm{ee}} [\rho^{\mathrm{e}}] + 
E_{\mathrm{exc}} [\rho^{\mathrm{e}}]  \\
+ T_{\mathrm{s}}^{\mathrm{p}} [\rho^{\mathrm{p}}] + 
V_{\mathrm{nuc}}^{\mathrm{p}} [\rho^{\mathrm{p}}] +
J^{\mathrm{pp}} [\rho^{\mathrm{p}}] + 
E_{\mathrm{pxc}} [\rho^{\mathrm{p}}]  
\\ + J^{\mathrm{ep}} [ \rho^{\mathrm{e}}, \rho^{\mathrm{p}}  ]
+ E_{\mathrm{epc}} [ \rho^{\mathrm{e}}, \rho^{\mathrm{p}}  ],
\end{multline}
where $T^{\mathrm{e}}_{\mathrm{s}} [\rho^{\mathrm{e}}]$ is the noninteracting kinetic energy of electrons,
$V_{\mathrm{nuc}}^{\mathrm{e}} [\rho^{\mathrm{e}}]$ is the interaction energy of electrons with classical nuclei, 
$J^{\mathrm{ee}} [\rho^{\mathrm{e}}]$ is the classical Coulomb interaction of electrons, and $E_{\mathrm{exc}} [\rho^{\mathrm{e}}]$ is the exchange--correlation energy, all well-known from standard DFT. The terms $T^{\mathrm{p}}_{\mathrm{s}} [\rho^{\mathrm{p}}]$,  $V_{\mathrm{nuc}}^{\mathrm{p}} [\rho^{\mathrm{p}}]$, 
$J^{\mathrm{pp}} [\rho^{\mathrm{p}}]$, and $E_{\mathrm{pxc}} [\rho^{\mathrm{p}}]$ are defined analogously for quantum protons and are functionals of the proton density $\rho^{\mathrm{p}} (\mathbf{r}^\mathrm{p})$. 
Note that, contrary to the corresponding electronic contributions, the expression for the kinetic energy $T^{\mathrm{p}}_{\mathrm{s}} [\rho^{\mathrm{p}}]$ also features the mass of a proton $m_{\mathrm{p}}$, while $V_{\mathrm{nuc}}^{\mathrm{p}} [\rho^{\mathrm{p}}]$ describes repulsive interactions of classical and quantum nuclei. The remaining terms $J^{\mathrm{ep}} [ \rho^{\mathrm{e}}, \rho^{\mathrm{p}} ]$ and $E_{\mathrm{epc}} [ \rho^{\mathrm{e}}, \rho^{\mathrm{p}} ]$ are the electron--proton Coulomb interaction and electron--proton correlation, respectively.

Assuming a closed-shell-type treatment of electrons and high-spin treatment of quantum protons (commonly used in NEO approaches~\cite{webb2002}), the minimization of the energy functional $E [ \rho^{\mathrm{e}}, \rho^{\mathrm{p}}  ]$ from Eq.~(\ref{eq:neo_energy}) with respect to electron $\rho^{\mathrm{e}}  (\mathbf{r}^{\mathrm{e}})$ and proton $\rho^{\mathrm{p}} (\mathbf{r}^{\mathrm{p}})$ densities leads to a coupled system of Kohn--Sham equations~\cite{chakraborty2008},
\begin{align}
\left[  -\frac{1}{2} \nabla^2_{\mathrm{e}} + \upsilon_{\mathrm{eff}}^{\mathrm{e}} [\rho^{\mathrm{e}}] (\mathbf{r}^{\mathrm{e}})  \right] \psi_i^{\mathrm{e}} (\mathbf{r}^{\mathrm{e}}) &= \varepsilon_i^{\mathrm{e}} \psi_i^{\mathrm{e}} (\mathbf{r}^{\mathrm{e}}),  \,\,\, i=1, \dots, N_{\mathrm{e}}/2, \label{eq:KS_eq_elec}\\
\left[  -\frac{1}{2m_{\mathrm{p}}} \nabla^2_{\mathrm{p}} + \upsilon_{\mathrm{eff}}^{\mathrm{p}} [\rho^{\mathrm{p}}] (\mathbf{r}^{\mathrm{p}})   \right] \psi_{i^{\prime}}^{\mathrm{p}} (\mathbf{r}^{\mathrm{p}}) &= \varepsilon_{i^{\prime}}^{\mathrm{p}} \psi_{i^{\prime}}^{\mathrm{p}} (\mathbf{r}^{\mathrm{p}}),  \,\,\, i^{\prime}=1, \dots, N_{\mathrm{p}} \label{eq:KS_eq_prot}.	
\end{align}
Here, $N_{\mathrm{e}}$ is the number of electrons and $N_{\mathrm{p}}$ is the number of quantum protons, $\psi_i^{\mathrm{e}} (\mathbf{r}^{\mathrm{e}})$ and $\psi_{i^{\prime}}^{\mathrm{p}} (\mathbf{r}^{\mathrm{p}})$ are Kohn--Sham orbitals of electrons and protons, respectively, whereas $\varepsilon_i^{\mathrm{e}}$ and $\varepsilon_{i^{\prime}}^{\mathrm{p}}$ are orbital energies. The indices $i$ and $i^{\prime}$ are introduced for electron and proton particles, respectively. 
The effective potentials are given as,
\begin{equation} \label{eq:pot_elec}
\upsilon_{\mathrm{eff}}^{\mathrm{e}} [\rho^{\mathrm{e}}] (\mathbf{r}^{\mathrm{e}}) = \upsilon_{\mathrm{nuc}}^{\mathrm{e}} (\mathbf{r}^{\mathrm{e}}) 
+ \upsilon_{\mathrm{Coul}}^{\mathrm{ee}} [\rho^{\mathrm{e}}] (\mathbf{r}^{\mathrm{e}})
+ \upsilon_{\mathrm{exc}} [\rho^{\mathrm{e}}] (\mathbf{r}^{\mathrm{e}}) \\
+ \upsilon_{\mathrm{Coul}}^{\mathrm{ep}} [\rho^{\mathrm{p}}] (\mathbf{r}^{\mathrm{e}})
+ \upsilon_{\mathrm{epc}}^{\mathrm{e}} [\rho^{\mathrm{e}}, \rho^{\mathrm{p}}] 
(\mathbf{r}^{\mathrm{e}}) ,
\end{equation}
\begin{equation}  \label{eq:pot_prot}
\upsilon_{\mathrm{eff}}^{\mathrm{p}} [\rho^{\mathrm{p}}] (\mathbf{r}^{\mathrm{p}}) = \upsilon_{\mathrm{nuc}}^{\mathrm{p}} (\mathbf{r}^{\mathrm{p}}) 
+ \upsilon_{\mathrm{Coul}}^{\mathrm{pp}} [\rho^{\mathrm{p}}] (\mathbf{r}^{\mathrm{p}})
+ \upsilon_{\mathrm{pxc}} [\rho^{\mathrm{p}}] (\mathbf{r}^{\mathrm{p}}) \\
+ \upsilon_{\mathrm{Coul}}^{\mathrm{ep}} [\rho^{\mathrm{e}}] (\mathbf{r}^{\mathrm{p}})
+ \upsilon_{\mathrm{epc}}^{\mathrm{p}} [\rho^{\mathrm{e}}, \rho^{\mathrm{p}}] (\mathbf{r}^{\mathrm{p}}).
\end{equation}
Here, $\upsilon_{\mathrm{nuc}}^{\mathrm{e}} (\mathbf{r}^{\mathrm{e}}) = \delta V_{\mathrm{nuc}}^{\mathrm{e}} [\rho^{\mathrm{e}}] / \delta \rho^{\mathrm{e}}$ and $\upsilon_{\mathrm{nuc}}^{\mathrm{p}} (\mathbf{r}^{\mathrm{p}})  = \delta V_{\mathrm{nuc}}^{\mathrm{p}} [\rho^{\mathrm{p}}] / \delta \rho^{\mathrm{p}}$
are nuclear potentials of electrons and quantum protons, respectively. The classical Coulomb potential $\upsilon_{\mathrm{Coul}}^{\mathrm{aa}} [\rho^{\mathrm{a}}] (\mathbf{r}^{\mathrm{a}})$ describing interactions of same-type particles is given as, 
\begin{equation}
\upsilon_{\mathrm{Coul}}^{\mathrm{aa}} [\rho^{\mathrm{a}}] (\mathbf{r}_1^{\mathrm{a}}) = \frac{\delta J^{\mathrm{aa}} [\rho^{\mathrm{a}}] }{\delta \rho^{\mathrm{a}} (\mathbf{r}_1^{\mathrm{a}})} = \int \frac{\rho^{\mathrm{a}} ( \mathbf{r}^{\mathrm{a}}_2 )   }{ | \mathbf{r}^{\mathrm{a}}_1  - \mathbf{r}^{\mathrm{a}}_2 | } \mathrm{d} \mathbf{r}^{\mathrm{a}}_2,
\end{equation}
whereas for different types of particles it reads,
\begin{equation}
\upsilon_{\mathrm{Coul}}^{\mathrm{ab}} [\rho^{\mathrm{b}}] (\mathbf{r}^{\mathrm{a}}) = \frac{\delta J^{\mathrm{ab}} [\rho^{\mathrm{a}}, \rho^{\mathrm{b}}] }{\delta \rho^{\mathrm{a}} (\mathbf{r}^{\mathrm{a}})} = - \int \frac{\rho^{\mathrm{b}} ( \mathbf{r}^{\mathrm{b}} )   }{ | \mathbf{r}^{\mathrm{b}}  - \mathbf{r}^{\mathrm{a}} | } \mathrm{d} \mathbf{r}^{\mathrm{b}}.
\end{equation}
The potentials $\upsilon_{\mathrm{exc}} [\rho^{\mathrm{e}}] (\mathbf{r}^{\mathrm{e}}) = \delta E_{\mathrm{exc}} [\rho^{\mathrm{e}}] / \delta \rho^{\mathrm{e}}$ and $\upsilon_{\mathrm{pxc}} [\rho^{\mathrm{p}}] (\mathbf{r}^{\mathrm{p}})  = \delta E_{\mathrm{pxc}} [\rho^{\mathrm{p}}] / \delta \rho^{\mathrm{p}}$ describe exchange and correlation of electrons and protons, respectively, while \linebreak $\upsilon_{\mathrm{epc}}^{\mathrm{e}} [\rho^{\mathrm{e}}, \rho^{\mathrm{p}}] (\mathbf{r}^{\mathrm{e}}) = \delta E_{\mathrm{epc}} [\rho^{\mathrm{e}}, \rho^{\mathrm{p}}] / \delta \rho^{\mathrm{e}}$  and $\upsilon_{\mathrm{epc}}^{\mathrm{p}} [\rho^{\mathrm{e}}, \rho^{\mathrm{p}}] (\mathbf{r}^{\mathrm{p}}) = \delta E_{\mathrm{epc}} [\rho^{\mathrm{e}}, \rho^{\mathrm{p}}] / \delta \rho^{\mathrm{p}}$ are the electron--proton correlation potentials. Additional superscripts ``e'' and ``p'' are used for the latter two terms to distinguish functional derivatives with respect to electron and proton densities, respectively.

As one can see from Eqs.~(\ref{eq:pot_elec}) and (\ref{eq:pot_prot}), in addition to standard DFT potentials and those defined in an analogous way, the unknown functional derivatives of $E_{\mathrm{pxc}} [ \rho^{\mathrm{p}}]$ and $E_{\mathrm{epc}} [\rho^{\mathrm{e}}, \rho^{\mathrm{p}}]$ are present. 
Due to the high localization of proton densities, the proton XC contribution $E_{\mathrm{pxc}} [ \rho^{\mathrm{p}}]$ is negligibly small and is not considered in NEO-DFT~\cite{sirjoosingh2011}. Instead, only the  proton--proton exchange is included on the Nuclear-Electronic Orbital Hartree--Fock (NEO-HF)~\cite{webb2002} level to avoid self-interaction. 
As for $E_{\mathrm{epc}} [\rho^{\mathrm{e}}, \rho^{\mathrm{p}}]$, several electron--proton correlation functionals were proposed in the recent years~\cite{yang2017,brorsen2017,brorsen2018,tao2019,Holzer2024,Hasecke2025}.

With all terms being defined, the solution of the system of equations from Eqs.~(\ref{eq:KS_eq_elec}) and (\ref{eq:KS_eq_prot}) can be achieved in, for example, a step-wise or simultaneous~\cite{Aodong2022} self-consistent field (SCF) optimization procedure providing electron $\rho^{\mathrm{e}} (\mathbf{r}^{\mathrm{e}})$ and proton densities $\rho^{\mathrm{p}} (\mathbf{r}^{\mathrm{p}})$ as well as the energy value $E [ \rho^{\mathrm{e}}, \rho^{\mathrm{p}}  ]$ with quantum effects of selected protons being incorporated in it.

\subsection{NEO-sDFT} \label{sec:neosdft}

To derive NEO-sDFT, we start from the  underlying idea of the sDFT approach~\cite{senatore1986,johnson1987,cortona1991,jacob2014,jacob2024}, and assume that the total molecular system can be partitioned into subsystems based on the electron density $\rho^{\mathrm{e}} (\mathbf{r}^{\mathrm{e}})$ such that
\begin{equation} \label{eq:dens_partitioning}
\rho^{\mathrm{e}} (\mathbf{r}^{\mathrm{e}}) = \sum_{I=1}^M \rho^{\mathrm{e}}_I (\mathbf{r}^{\mathrm{e}}),
\end{equation} 
where $M$ is the number of subsystems and $\rho^{\mathrm{e}}_I (\mathbf{r}^{\mathrm{e}})$ are subsystem electron densities defined via KS-like orbitals $\psi^{\mathrm{e}}_{i_I}$ of corresponding subsystem $I$,
\begin{equation}
\rho^{\mathrm{e}}_I (\mathbf{r}^{\mathrm{e}}) = 2 \sum_{i_I=1}^{N^I_e / 2} | \psi^{\mathrm{e}}_{i_I} (\mathbf{r}^{\mathrm{e}}) |^2.
\end{equation}
These orbitals are further expressed as linear combinations of atomic orbitals (AOs), which, in practical computations, are often centered only at the nuclei of subsystem $I$, i.e., we assume that a monomer basis set is employed. In this case, this density partitioning uniquely defines the partitioning of the system of nuclei as well. We further adopt an analogous expression for the proton density $\rho^{\mathrm{p}} (\mathbf{r}^{\mathrm{p}})$ splitting it into $M$ subsystem contributions,
\begin{equation} \label{eq:dens_parti}
\rho^{\mathrm{p}} (\mathbf{r}^{\mathrm{p}}) = \sum_{I=1}^M \rho^{\mathrm{p}}_I (\mathbf{r}^{\mathrm{p}}).
\end{equation}
If subsystem $K$ does not contain quantum protons, then the corresponding contribution $\rho^{\mathrm{p}}_K (\mathbf{r}^{\mathrm{p}}) $ is simply equal to zero. 
With this common partitioning scheme for both  densities being introduced, it is now possible to re-write energy terms from Eq.~(\ref{eq:neo_energy}) as,
\begin{equation} \label{eq:kin_partition}
T_{\mathrm{s}}^{\mathrm{e}} [ \rho^{\mathrm{e}}] = \sum_{I=1}^M T_{\mathrm{s}}^{\mathrm{e}} [ \rho_I^{\mathrm{e}}] + T_{\mathrm{s}}^{\mathrm{e, nad}} [\{ \rho_I^{\mathrm{e}} \}],
\end{equation}
\begin{equation} \label{eq:nuc_partition}
V_{\mathrm{nuc}}^{\mathrm{e}} [\rho^{\mathrm{e}}]  = \sum_{I=1}^M V_{\mathrm{nuc}}^{\mathrm{e},I} [\rho_I^{\mathrm{e}}] + \sum_{I=1}^M \sum_{J \ne I}^M V_{\mathrm{nuc}}^{\mathrm{e},J} [\rho_I^{\mathrm{e}}],
\end{equation}
\begin{equation}
J^{\mathrm{ee}} [\rho^{\mathrm{e}}]  = \sum_{I=1}^M J^{\mathrm{ee}} [\rho_I^{\mathrm{e}}] + \sum_{I=1}^M \sum_{J > I}^M J^{\mathrm{ee}} [\rho_I^{\mathrm{e}}, \rho_J^{\mathrm{e}}],
\end{equation}
\begin{equation} \label{eq:exc_partition}
E_{\mathrm{exc}} [\rho^{\mathrm{e}}]  = \sum_{I=1}^M E_{\mathrm{exc}} [\rho_I^{\mathrm{e}}] + E_{\mathrm{exc}}^{\mathrm{nad}} [\{ \rho_I^{\mathrm{e}} \}],
\end{equation}
\begin{equation} \label{eq:Jep_partition}
J^{\mathrm{ep}} [ \rho^{\mathrm{e}}, \rho^{\mathrm{p}}  ] = \sum_{I=1}^M J^{\mathrm{ep}} [ \rho_I^{\mathrm{e}}, \rho_I^{\mathrm{p}}  ] + \sum_{I=1}^M \sum_{J \ne I}^M J^{\mathrm{ep}} [ \rho_I^{\mathrm{e}}, \rho_J^{\mathrm{p}}  ],  
\end{equation}
\begin{equation} \label{eq:epc_partition}
E_{\mathrm{epc}} [ \rho^{\mathrm{e}}, \rho^{\mathrm{p}}  ] = \sum_{I=1}^M E_{\mathrm{epc}} [ \rho_I^{\mathrm{e}}, \rho_I^{\mathrm{p}} ] + E_{\mathrm{epc}}^{\mathrm{nad}} [ \{ \rho_I^{\mathrm{e}} \}, \{ \rho_J^{\mathrm{p}} \} ].
\end{equation}
In these expressions, the first term on the right-hand side is the sum of purely subsystem contributions, whereas the second term describes non-additive contributions coming from interactions between different subsystems. The curly braces notations $\{ \rho^{\mathrm{e}}_I \}$ are used to denote a set of densities $\rho^{\mathrm{e}}_I$ with $I = 1, \dots, M$. All terms from Eqs.~(\ref{eq:kin_partition})--(\ref{eq:exc_partition}) are known from electronic sDFT~\cite{senatore1986,johnson1987,cortona1991,jacob2014,jacob2024}, whereas  
the proton energies
$T_{\mathrm{s}}^{\mathrm{p}} [\rho^{\mathrm{p}}]$, 
$V_{\mathrm{nuc}}^{\mathrm{p}} [\rho^{\mathrm{p}}]$, $J^{\mathrm{pp}} [\rho^{\mathrm{p}}]$, and $
E_{\mathrm{pxc}} [\rho^{\mathrm{p}}]$ are expressed in full analogy
to their electronic counterparts and are not explicitly given here. 
As long as all energy functionals from Eqs.~(\ref{eq:nuc_partition})--(\ref{eq:epc_partition}) 
are purely-density dependent, it is straightforward to compute the corresponding non-additive contributions, i.e., all subsystem and total densities are available for functional evaluation. 
However, the non-interacting kinetic energies $T_{\mathrm{s}}^{\mathrm{e}} [ \rho^{\mathrm{e}} ]$ and $T_{\mathrm{s}}^{\mathrm{p}} [ \rho^{\mathrm{p}} ]$ are expressed in terms of orbitals and are not known as explicit functionals of densities. Therefore, the related non-additive terms 
$T_{\mathrm{s}}^{\mathrm{e, nad}} [\{ \rho_I^{\mathrm{e}} \}]$ and 
$T_{\mathrm{s}}^{\mathrm{p, nad}} [\{ \rho_I^{\mathrm{p}} \}]$ have to be approximated. A plethora of different explicit functionals of density $T_{\mathrm{s}}^{\mathrm{e, nad}} [\{ \rho_I^{\mathrm{e}} \}]$ was proposed in sDFT for this purpose (for reviews, see Refs.~\citenum{jacob2014,jacob2024}), which can be directly applied in our NEO-sDFT approach as well. 
However, the approximation of the non-additive kinetic energy by such explicit density functionals constitutes the main error source in sDFT (for example, see Ref.~\citenum{arti2015}) and limits its applicability to weakly interacting subsystems, unless more advanced techniques, such as projection-based embedding~\cite{manby2012,Tamukong2014} or potential reconstruction~\cite{Roncero2008,fux2010,manby_miller2010,huang2011} are employed. It is also anticipated to be the main error source in NEO-sDFT. Contrary to that, $T_{\mathrm{s}}^{\mathrm{p, nad}} [\{ \rho_I^{\mathrm{p}} \}]$ is negligible due to small overlaps of proton densities and, therefore, is not considered in this work. For the same reason, the non-additive proton exchange--correlation contributions $E_{\mathrm{pxc}}^{\mathrm{nad}} [\{ \rho_I^{\mathrm{p}} \}]$, the non-additive electron--proton correlation $E_{\mathrm{epc}}^{\mathrm{nad}} [\{ \rho_I^{\mathrm{e}} \}, \{ \rho_J^{\mathrm{p}} \}]$, the intermolecular proton--proton exchange, and potentials related to them are not computed (for further justifications of this, see Section~\nameref{sec:water_dimers}). However, we keep these terms in the following mathematical expressions for the sake of completeness and generality.

Minimizing the total energy functional $E [ \rho^{\mathrm{e}}, \rho^{\mathrm{p}}  ]$ with respect to densities $\rho_K^{\mathrm{e}}  (\mathbf{r}^{\mathrm{e}})  $ or $\rho_K^{\mathrm{p}} (\mathbf{r}^{\mathrm{p}}) $ of subsystem $K$ and keeping all other densities frozen, the Kohn--Sham equations with constrained electron and proton densities can be derived in close analogy to sDFT~\cite{senatore1986,johnson1987,cortona1991,jacob2014,jacob2024} and multicomponent density functional theory embedding~\cite{culpitt2016}. The final expressions read,
\begin{multline} \label{eq:KSCED_eq_elec_subsystem}
\left[  -\frac{1}{2} \nabla^2_{\mathrm{e}} + \upsilon_{\mathrm{eff}}^{\mathrm{e}, K}  [\rho_K^{\mathrm{e}}, \rho_K^{\mathrm{p}}] (\mathbf{r}^{\mathrm{e}}) + \upsilon_{\mathrm{emb}}^{\mathrm{e}, K} [\rho^{\mathrm{e}}, \rho^{\mathrm{p}}] (\mathbf{r}^{\mathrm{e}})  \right] \psi_{i_K}^{\mathrm{e}} (\mathbf{r}^{\mathrm{e}}) \\ = \varepsilon_{i_K}^{\mathrm{e}} \psi_{i_K}^{\mathrm{e}} (\mathbf{r}^{\mathrm{e}}),  \,\,\, {i_K}=1, \dots, N^K_{\mathrm{e}}/2 ,
\end{multline}
\begin{multline} \label{eq:KSCED_eq_prot_subsystem}
\left[  -\frac{1}{2m_{\mathrm{p}}} \nabla^2_{\mathrm{p}} + \upsilon_{\mathrm{eff}}^{\mathrm{p}, K} [\rho_K^{\mathrm{e}}, \rho_K^{\mathrm{p}}] (\mathbf{r}^{\mathrm{p}}) + \upsilon_{\mathrm{emb}}^{\mathrm{p}, K} [\rho^{\mathrm{e}}, \rho^{\mathrm{p}}] (\mathbf{r}^{\mathrm{p}})  \right] \psi_{i_K^{\prime}}^{\mathrm{p}} (\mathbf{r}^{\mathrm{p}}) \\ = \varepsilon_{i_K^{\prime}}^{\mathrm{p}} \psi_{i_K^{\prime}}^{\mathrm{p}} (\mathbf{r}^{\mathrm{p}}),  \,\,\, i_K^{\prime}=1, \dots, N^K_{\mathrm{p}} .	
\end{multline}
Here, the numbers of particles $N^K_{\mathrm{e}}$ and $N^K_{\mathrm{p}}$, KS-like orbitals $\psi_{i_K}^{\mathrm{e}}$ and $\psi_{i^{\prime}_K}^{\mathrm{p}}$, orbital energies $\varepsilon_{i_K}^{\mathrm{e}}$ and $\varepsilon_{i_K^{\prime}}^{\mathrm{p}}$, and 
effective potentials $\upsilon_{\mathrm{eff}}^{\mathrm{e}, K}$ and $\upsilon_{\mathrm{eff}}^{\mathrm{p}, K}$ are all analogous to those from 
Eqs.~(\ref{eq:KS_eq_elec})--(\ref{eq:pot_prot}) of NEO-DFT, but defined for subsystem $K$. 
The new embedding potentials $\upsilon_{\mathrm{emb}}^{\mathrm{e}, K} [\rho^{\mathrm{e}}, \rho^{\mathrm{p}}] (\mathbf{r}^{\mathrm{e}})$ and $\upsilon_{\mathrm{emb}}^{\mathrm{p}, K} [\rho^{\mathrm{e}}, \rho^{\mathrm{p}}] (\mathbf{r}^{\mathrm{p}})$ account for the interaction between subsystems and are equal to the sums of functional derivatives of the non-additive energy terms discussed above,
\begin{multline}
\upsilon_{\mathrm{emb}}^{\mathrm{e}, K} [\rho^{\mathrm{e}}, \rho^{\mathrm{p}}] (\mathbf{r}^{\mathrm{e}}) = \sum_{I \ne K}^M \upsilon_{\mathrm{nuc}}^{\mathrm{e}, I} (\mathbf{r}^{\mathrm{e}})
+ \upsilon_{\mathrm{Coul}}^{\mathrm{ee}} [\rho^{\mathrm{e}} - \rho_K^{\mathrm{e}}] (\mathbf{r}^{\mathrm{e}}) 
+ \upsilon_{\mathrm{exc}}^{\mathrm{nad}} [\rho^{\mathrm{e}}, \rho_K^{\mathrm{e}}] (\mathbf{r}^{\mathrm{e}}) \\
+ \upsilon_{\mathrm{kin}}^{\mathrm{e, nad}} [\rho^{\mathrm{e}}, \rho_K^{\mathrm{e}}] (\mathbf{r}^{\mathrm{e}})
+ \upsilon_{\mathrm{Coul}}^{\mathrm{ep}} [\rho^{\mathrm{p}} - \rho_K^{\mathrm{p}}] (\mathbf{r}^{\mathrm{e}})
+ \upsilon_{\mathrm{epc}}^{\mathrm{e, nad}}  [ \rho^{\mathrm{e}}, \rho_K^{\mathrm{e}}, \rho^{\mathrm{p}}, \rho_K^{\mathrm{p}} ]  ,
\end{multline}
\begin{multline}
\upsilon_{\mathrm{emb}}^{\mathrm{p}, K} [\rho^{\mathrm{e}}, \rho^{\mathrm{p}}] (\mathbf{r}^{\mathrm{p}}) =  \sum_{I \ne K}^M \upsilon_{\mathrm{nuc}}^{\mathrm{p}, I} (\mathbf{r}^{\mathrm{p}})
+ \upsilon_{\mathrm{Coul}}^{\mathrm{pp}} [\rho^{\mathrm{p}} - \rho_K^{\mathrm{p}}] (\mathbf{r}^{\mathrm{p}}) 
+ \upsilon_{\mathrm{pxc}}^{\mathrm{nad}} [\rho^{\mathrm{p}}, \rho_K^{\mathrm{p}}] (\mathbf{r}^{\mathrm{p}}) \\
+ \upsilon_{\mathrm{kin}}^{\mathrm{p, nad}} [\rho^{\mathrm{p}}, \rho_K^{\mathrm{p}}] (\mathbf{r}^{\mathrm{p}})
+ \upsilon_{\mathrm{Coul}}^{\mathrm{ep}} [\rho^{\mathrm{e}} - \rho_K^{\mathrm{e}}] (\mathbf{r}^{\mathrm{p}})
+ \upsilon_{\mathrm{epc}}^{\mathrm{p, nad}}  [ \rho^{\mathrm{e}}, \rho_K^{\mathrm{e}}, \rho^{\mathrm{p}}, \rho_K^{\mathrm{p}} ]                .
\end{multline}
Eqs.~(\ref{eq:KSCED_eq_elec_subsystem}) and (\ref{eq:KSCED_eq_prot_subsystem}) can be solved by means of freeze-and-thaw cycles~\cite{weso1996}, often employed in electronic sDFT, relaxing electron and proton densities of each individual subsystem one by one while keeping others frozen. When this procedure is repeated until full convergence of all electron and proton densities, the method can be referred to as NEO-sDFT. In case of a partial optimization, we will call it Nuclear-Electronic Orbital Frozen-Density Embedding (NEO-FDE) similar to the existing Frozen-Density Embedding (FDE) approach~\cite{wesolowski1993}.

This formulation of embedding theory differs from that presented in Ref.~\citenum{culpitt2016} in a few key aspects. First, in our derivations a monomer basis set is employed and the system of classical nuclei is therefore uniquely partitioned into subsystems alongside the electron density.
This results in non-zero contributions from the electron--nuclei interaction $V_{\mathrm{nuc}}^{\mathrm{e},J} [\rho_I^{\mathrm{e}}]$ between subsystems.
Secondly, we introduced a partitioning of the proton density with Eq.~(\ref{eq:dens_parti}). As a consequence, the new proton embedding potential $\upsilon_{\mathrm{emb}}^{\mathrm{p}, K} [\rho^{\mathrm{e}}, \rho^{\mathrm{p}}] (\mathbf{r}^{\mathrm{p}})$ naturally appears allowing us to consider quantum protons not only in the active but also in the environment subsystems.
In other words, while the theory from Ref.~\citenum{culpitt2016} can be classified as (NEO-DFT)-in-DFT embedding, our approach additionally enables (NEO-DFT)-in-(NEO-DFT) embedding computations. 
This should not be understood as a limitation of our approach, as the presented program implementation is very flexible and allows one to consider any number of quantum protons (or none) in different subsystems.
Finally, no orthogonality constraints are enforced on subsystem KS-like orbitals of electrons or protons leading to the need to approximate the non-additive kinetic energy contribution $T_{\mathrm{s}}^{\mathrm{e, nad}} [\{ \rho_I^{\mathrm{e}} \}]$. This is the most crucial difference as it makes it possible to compute large molecular systems including thousands of atoms. At the same time, this restricts our approach to molecular systems composed of rather weakly interacting subsystems.

\section{Computational Details} \label{sec:comput_details}

\begin{figure}
	\centering
	\includegraphics[width=0.8\textwidth]{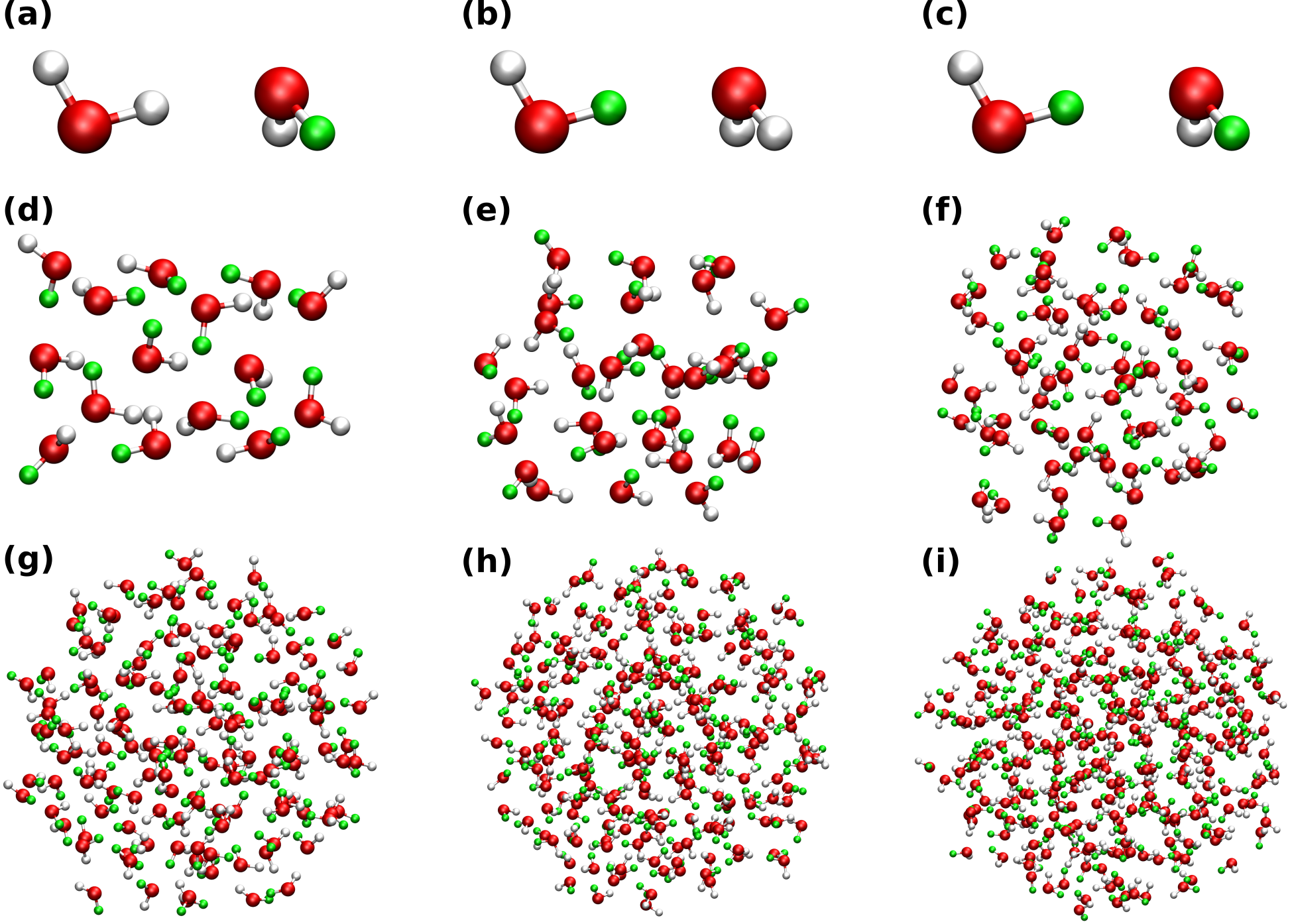}
	\caption{Some of the water clusters (H$_2$O)$_M$ employed in this work. The number of water molecules $M$ is equal to (a)--(c) 2, (d) 15, (e) 30, (f) 70, (g) 145, (h) 237 and (i) 303. Classical oxygen atoms are shown in red, while classical hydrogen atoms are in white. Hydrogen nuclei, which are considered quantum mechanically, are shown in green.}
	\label{fig:structures}
\end{figure}

Some of the molecular structures computed in this work are presented in Figure~\ref{fig:structures}. Displaced structures of a water dimer, as shown in Figures~\ref{fig:structures} (a)--(c), were taken from Ref.~\citenum{Eitelhuber2025}. Small water clusters (H$_2$O)$_M$ with $M=3, 4, 5, 6, 10, 12, 15$ and 30, the latter two are shown in Figures~\ref{fig:structures} (d) and (e), respectively, were taken from Ref.~\citenum{Bandow2006}, whereas larger models with $M=70$, 145, 237, and 303 from Figure~\ref{fig:structures} (f)--(i), respectively, were extracted from the accompanying data~\cite{water_clusters} to the \textsc{Ergo} software~\cite{Ergo2018}. Classical protons were subsequently replaced with their quantum counterparts without further structure optimization. For testing purposes, we considered different numbers of quantum protons in the dimeric models, whereas in larger water clusters one proton per water molecule was considered quantum mechanically, unless stated otherwise. 
Specific choices of quantum protons in water dimers and their consequences for the accuracy are discussed in Sec.~\nameref{sec:water_dimers}. In larger clusters, the first proton appearing in each subsystem molecular structure was chosen without considering its participation in hydrogen bonding. This rather arbitrary assignment is sufficient for assessing the computational efficiency of our approach. We note that in practical computations, the quantum protons should be selected based on the chemical problem at hand. In this regard, the presented approach is highly flexible and allows one to assign any number of quantum protons to any subsystems. However, to avoid contamination of proton densities with rotations and translations, at least two nuclei per subsystem must remain classical~\cite{neo_review2020}.

All single point computations were performed using a locally-modified version of the  \textsc{Serenity} software package~\cite{unsleber2018,niemeyer2023,serenity_zenodo}, where the  NEO-sDFT approach was implemented. For comparison purposes, DFT, sDFT~\cite{senatore1986,johnson1987,cortona1991,jacob2014,jacob2024}, and NEO-DFT~\cite{pak2007,chakraborty2008,chakraborty2009,sirjoosingh2011,sirjoosingh2012} computations were also performed. The reliability of the NEO-DFT program implementation in \textsc{Serenity} was demonstrated previously in Ref.~\citenum{artiukhin2026}.
In computations of energies and densities, the augmented correlation--consistent polarized valence triple-$\zeta$ aug-cc-pVTZ
basis set~\cite{Dunning1989} was used, whereas wall time measurements were performed using the valence triple-$\zeta$ def2-TZVPP basis~\cite{Weigend2005} with two sets of polarization functions for a better comparison with the results presented in Ref.~\citenum{unsleber2018}. 
The PW91 exchange--correlation (XC) functional~\cite{perdew1991,perdew1992} was employed in all computations. The integration grids were constructed using the Lebedev angular~\cite{Lebedev1999} and
Ahlrichs radial~\cite{Treutler1995,Krack1998} schemes with Stratmann--Scuseria--Frisch partitioning~\cite{Stratmann1996}.
For accurate computations of interaction energies presented in Section~\nameref{sec:water_dimers}, we employed very fine integration grids.
During self-consistent field iterations, grids with internal \textsc{Serenity} accuracy parameter 5 corresponding to about 40--50 radial shells and 50--434 angular points per shell were used. A larger grid (``accuracy 6'') with about 45--55 radial shells and 110--590 angular grid points per shell was used for the final energy evaluation.
Tight SCF convergence thresholds of 1.0$\times$10$^{-8}$, 1.0$\times$10$^{-8}$, and 1.0$\times$10$^{-7}$ a.u.\ were set for the changes in the total energy, electronic and protonic density matrix elements, and the direct inversion in the iterative subspace (DIIS) error, respectively. To demonstrate computational scaling of our approach in Section~\nameref{sec:large_scale}, looser default \textsc{Serenity} settings were used for integration grids and convergence thresholds (for more details, see Refs.\citenum{unsleber2018,niemeyer2023,serenity_zenodo}).

NEO-DFT, NEO-sDFT, and NEO-FDE computations were carried out using the proton PB4-D basis set~\cite{Yu2020} and 
the electron--proton correlation functional epc17-2~\cite{brorsen2017}. In all these cases, the SCF procedure was initiated with a core guess for quantum protons by Mata and co-workers~\cite{Hasecke2023} and performed using the simultaneous protocol~\cite{Aodong2022}. Density fitting techniques were not used for either electron or proton densities.

In sDFT and NEO-sDFT/FDE computations, the non-additive kinetic energy functional \linebreak PW91k~\cite{lembarki1994} was employed. The density partitioning was performed such that each subsystem contained a single water molecule, unless stated otherwise.
NEO-sDFT calculations for water dimers as presented in Section~\nameref{sec:water_dimers} were carried out fully self-consistently using freeze-and-thaw cycles~\cite{weso1996} until the sum of absolute element-wise differences in density matrices between subsequent iterations was below 1.0$\times 10^{-6}$~a.u. In NEO-FDE and FDE calculations of water clusters in Section~\nameref{sec:large_scale}, a fixed number of freeze-and-thaw cycles equal to three was used. The wall times were measured while performing parallel computations on a single Intel Xeon Skylake 6130 @ 2.1 GHz compute node with 32 CPU cores in total.

\section{Results and Discussion} \label{sec:results}

To validate our NEO-sDFT approach, we first assess its accuracy in comparison with NEO-DFT in Section~\nameref{sec:water_dimers}. Subsequently, we demonstrate its efficiency and favorable computational scaling in Section~\nameref{sec:large_scale} by computing molecular systems of increasing size up to 909 atoms in total with 303 protons treated quantum mechanically.

\subsection{NEO-sDFT Computations of Energies and Densities} \label{sec:water_dimers}

As mentioned above, the NEO-sDFT approach can be viewed as an approximation to NEO-DFT due to the need to compute the electronic non-additive kinetic energy 
$T_{\mathrm{s}}^{\mathrm{e, nad}} [\{ \rho_I^{\mathrm{e}} \}]$ with explicit density functionals. We assess the introduced error by computing the interaction energy of water dimers at different intermolecular displacements. This error should, in principle, be comparable in magnitude to that in the electronic sDFT method, where the same approximation for $T_{\mathrm{s}}^{\mathrm{e, nad}} [\{ \rho_I^{\mathrm{e}} \}]$ is invoked. For this reason, we also present sDFT and DFT interaction energies. Another motivation for performing these test computations is to verify that the 
neglected NEO-sDFT energy contributions such as the proton non-additive kinetic energy $T_{\mathrm{s}}^{\mathrm{p, nad}} [\{ \rho_I^{\mathrm{p}} \}]$ and the non-additive electron--proton correlation $E_{\mathrm{epc}}^{\mathrm{nad}} [\{ \rho_I^{\mathrm{p}} \}]$
are indeed very small and do not affect the quality of results.

\begin{figure}
	\centering
	\includegraphics[width=0.8\textwidth]{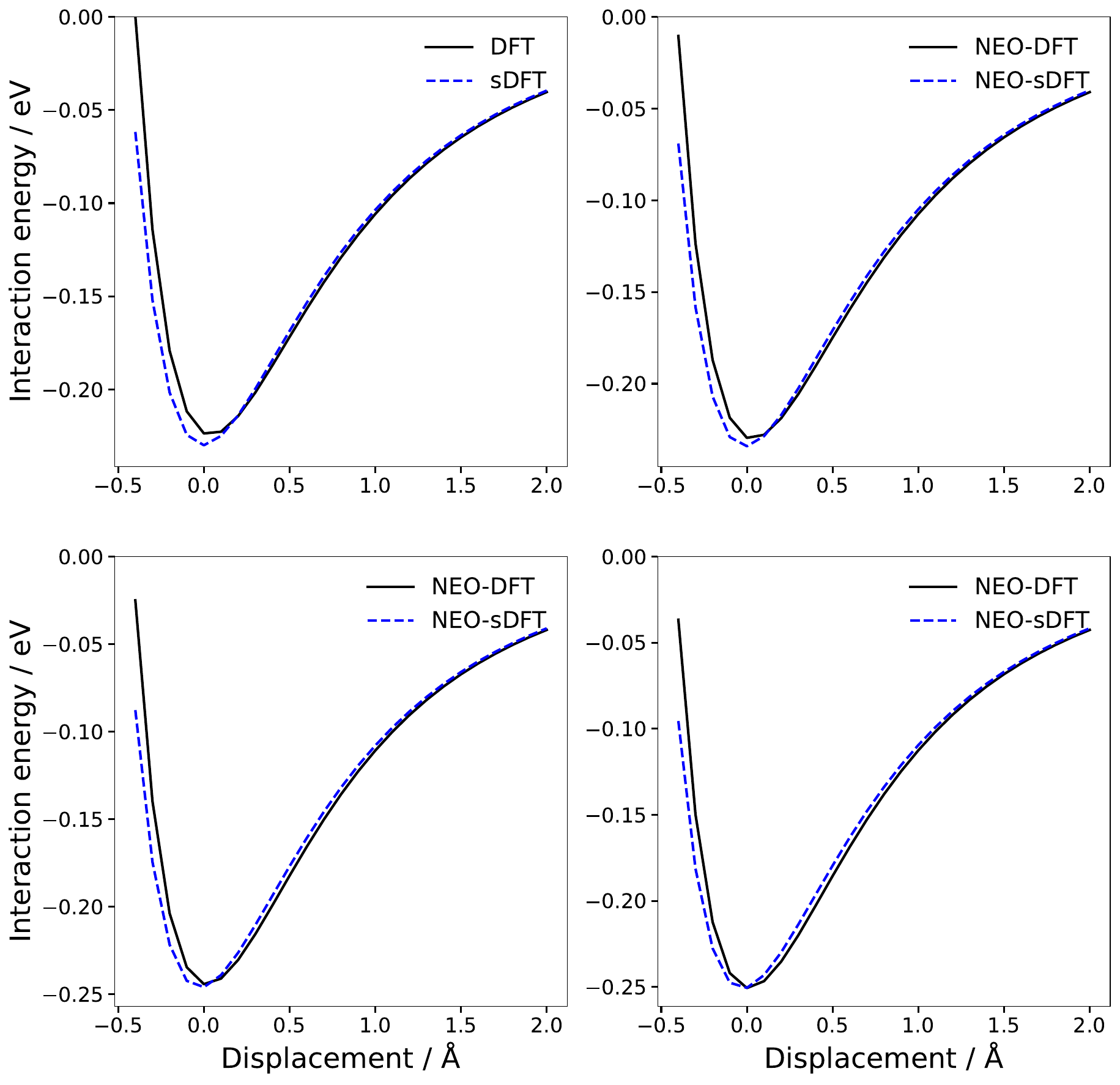}
	\caption{Interaction energies of water dimers HOH$\cdots$OHH (top left),  HOH$\cdots$OH\textbf{H} (top right), HO\textbf{H}$\cdots$OHH (bottom left), and HO\textbf{H}$\cdots$OH\textbf{H} (bottom right). See the main text for more details.}
	\label{fig:PESs}
\end{figure}

We use bold fonts to denote quantum protons and consider four different scenarios: 
\begin{enumerate}
	\item HOH$\cdots$OHH, which does not contain quantum protons; 
	\item HOH$\cdots$OH\textbf{H} with one quantum proton, which does not form a hydrogen bond between molecules/subsystems;
	\item HO\textbf{H}$\cdots$OHH with one quantum proton, forming a hydrogen bond;
	\item HO\textbf{H}$\cdots$OH\textbf{H} containing two quantum protons.
\end{enumerate}
DFT and sDFT computations of the first model HOH$\cdots$OHH provide an estimate for the error in $T_{\mathrm{s}}^{\mathrm{e, nad}} [\{ \rho_I^{\mathrm{e}} \}]$. NEO-DFT and NEO-sDFT computations for the second and third models additionally demonstrate the influence of the non-additive part of the electron--proton correlation $E_{\mathrm{epc}}^{\mathrm{nad}} [\{ \rho_I^{\mathrm{e}} \}, \{ \rho_J^{\mathrm{p}} \} ]$, when the overlap of electron and proton densities belonging to different subsystems is small and large, respectively. Finally, in the fourth model, one proton per water molecule is considered. As a result, the total proton density $\rho^{\mathrm{p}} (\mathbf{r}^{\mathrm{p}})$ is partitioned into two subsystem contributions $\rho_1^{\mathrm{p}} (\mathbf{r}^{\mathrm{p}})$ and $\rho_2^{\mathrm{p}} (\mathbf{r}^{\mathrm{p}})$ giving rise to the non-additive kinetic energy of protons $T_{\mathrm{s}}^{\mathrm{p, nad}} [\{ \rho_I^{\mathrm{p}} \}]$, which is neglected in our computations. 
We note, however, that other error sources are also present such as those coming from the use of a monomer basis set, numerical integration or neglect of the basis set superposition error. Therefore, the presented tests provide only approximate errors. 
In particular, since the reference NEO-DFT calculations employ supermolecular basis sets, the energy deviations presented below include the error due to the monomer basis set. The use of monomer basis sets is known to limit the accuracy of sDFT, especially for charge-transfer phenomena~\cite{Laricchia2013,Roy2025}, and supermolecular basis sets are required when the accuracy of non-additive functionals is to be assessed in isolation (for examples, see Refs.~\citenum{Wesolowski1997,arti2015}). For quantum protons, in contrast, monomer basis sets are expected to be sufficient due to the high localization of proton densities. Therefore, NEO-sDFT can be expected to inherit the basis-set requirements of electronic sDFT. For a detailed discussion on other error sources when computing interaction energy curves with sDFT, we refer the interested reader to Ref.~\citenum{schluens2015}.
A more thorough error assessment of NEO-sDFT will be conducted elsewhere. The results for all four scenarios are shown in Figure~\ref{fig:PESs}.

As one can see from Figure~\ref{fig:PESs}~(top left), interaction energies from sDFT are very similar to those from DFT. The energy curves almost coincide for intermolecular displacement larger than 0.2~\AA. The deviation becomes larger for smaller intermolecular distances with the error of about 6~meV for the equilibrium separation of 0.0~{\AA}. For the shortest displacement of $-$0.4~{\AA}, the deviation reaches about 62~meV. Moving to the second scenario HOH$\cdots$OH\textbf{H} from Figure~\ref{fig:PESs}~(top right) and comparing the performance of NEO-sDFT and NEO-DFT approaches, one finds slightly smaller errors of 5~meV and 59~meV for displacements of 0.0~{\AA} and $-$0.4~{\AA}, respectively. It is interesting to note that even smaller errors at 0.0~{\AA} are computed for the third HOH$\cdots$OH\textbf{H} and fourth HO\textbf{H}$\cdots$OH\textbf{H} models, as seen from Figures~\ref{fig:PESs}~(bottom left) and (bottom right), being equal to about 2~meV and 1~meV, respectively. However, this trend does not hold for the errors computed at the displacement of $-$0.4~{\AA}. In this case, the error is equal to 63~meV for HOH$\cdots$OH\textbf{H} and 59~meV for HO\textbf{H}$\cdots$OH\textbf{H}. 

\begin{figure}
	\centering
	\includegraphics[width=0.7\textwidth]{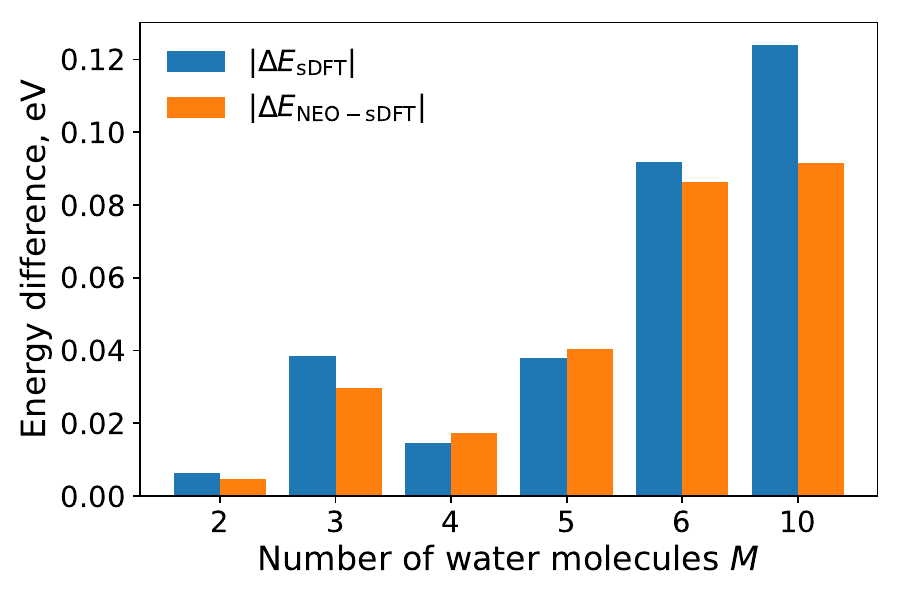}
	\caption{Absolute differences in total energies between DFT and sDFT, $|\Delta E_{\text{sDFT}}| = |E_{\text{DFT}} - E_{\text{sDFT}}|$, as well as between NEO-DFT and NEO-sDFT, $|\Delta E_{\text{NEO-sDFT}}| = |E_{\text{NEO-DFT}} - E_{\text{NEO-sDFT}}|$, for clusters containing $M = 2$, 3, 4, 5, 6, and 10 water molecules. In the NEO computations, one proton per supersystem was treated quantum mechanically.}
	\label{fig:larger_clusters_energies}
\end{figure}

Absolute differences between DFT and sDFT as well as between NEO-DFT and NEO-sDFT total energies for clusters containing up to ten water molecules are further shown in Figure~\ref{fig:larger_clusters_energies}. In NEO-DFT and NEO-sDFT computations, one proton per supersystem was treated quantum mechanically. As can be seen, sDFT and NEO-sDFT perform similarly, yielding comparable deviations from the reference results. This again points to the fact that approximations in sDFT, such as the electronic non-additive kinetic energy $T_{\mathrm{s}}^{\mathrm{e,nad}}[\{\rho_I^{\mathrm{e}}\}]$ and the use of a monomer basis set for electrons, constitute the major error sources of NEO-sDFT as well. The errors tend to grow with the system size and reach about 0.1~eV for the largest cluster considered. NEO-sDFT produces slightly better results than sDFT for molecular systems with $M=2$, 3, 6, and 10 waters. Summarizing the results for energies of water dimers and larger clusters, we conclude that NEO-sDFT is a reliable approach and the influence of the neglected energy contributions $T_{\mathrm{s}}^{\mathrm{p,nad}}[\{\rho_I^{\mathrm{p}}\}]$ and $E_{\mathrm{epc}}^{\mathrm{nad}}[\{\rho_I^{\mathrm{e}}\},\{\rho_I^{\mathrm{p}}\}]$ is rather small. Moreover, due to favorable error cancellation effects, NEO-sDFT can produce better results than sDFT.

\begin{figure}
	\centering
	\includegraphics[width=0.99\textwidth]{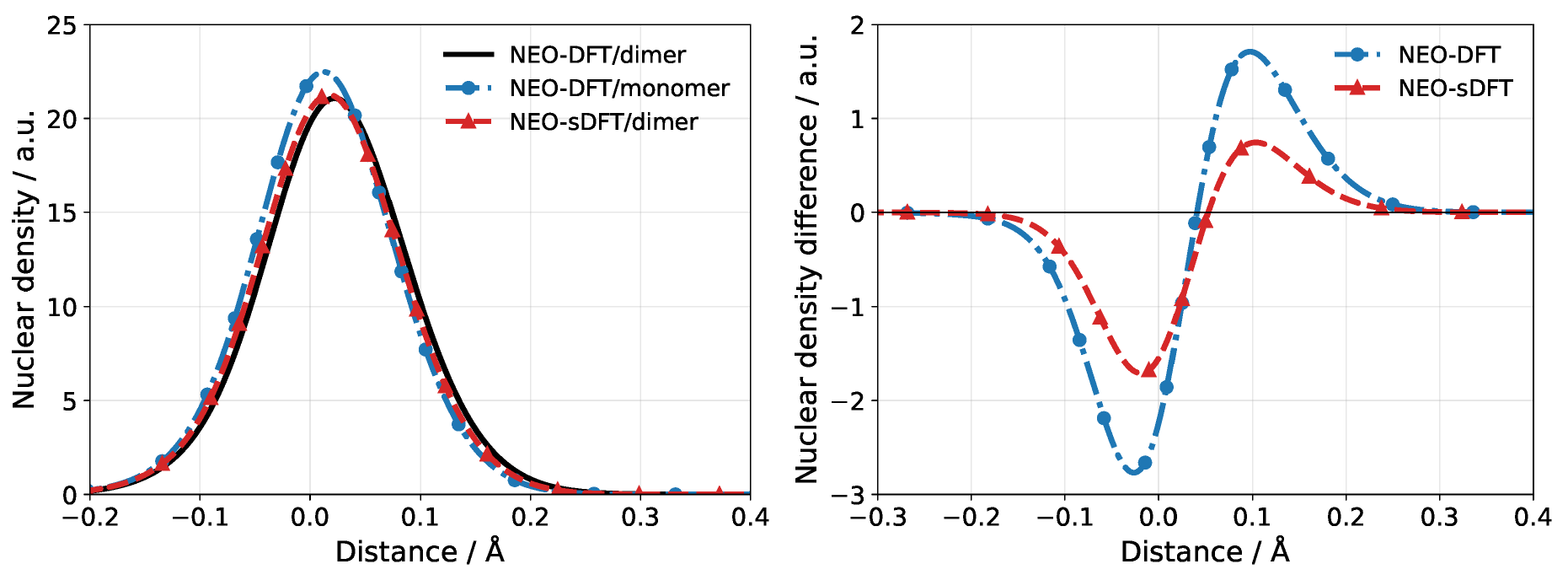}
	\caption{Proton density distributions (left) and proton density differences (right) computed for the water dimer HO\textbf{H}$\cdots$OHH and isolated monomer HO\textbf{H} along the intermolecular hydrogen bond \textbf{H}$\cdots$O. The difference is calculated as $\rho_{X}^{\text{dimer}} - \rho_{\text{NEO-DFT}}^{\text{monomer}}$, where $X$ is NEO-DFT or NEO-sDFT.}
	\label{fig:dens_polarization}
\end{figure}

With the energy values being analyzed, we also demonstrate that NEO-sDFT can account for proton density polarization effects induced by the environment subsystems. To that end, we consider the model HO\textbf{H}$\cdots$OHH and compute the density distribution of the quantum proton along the hydrogen bond \textbf{H}$\cdots$O. Computations are performed for the water dimer at the equidistant separation with NEO-sDFT and NEO-DFT, where density polarization effects are in play, as well as for the isolated subsystem HO\textbf{H} with NEO-DFT. Resulting density plots are shown in Figure~\ref{fig:dens_polarization} (left), whereas nuclear density differences are demonstrated in Figure~\ref{fig:dens_polarization} (right). As one can see from Figure~\ref{fig:dens_polarization} (left), NEO-sDFT and NEO-DFT produce very similar proton density distributions in the dimer, which are slightly more delocalized and shifted towards the oxygen atom compared to the proton density of the isolated monomer. The difference plot from Figure~\ref{fig:dens_polarization} (right) further demonstrates that NEO-sDFT produces less polarized density than the reference NEO-DFT approach. This points to a high sensitivity of the density polarization to the approximations inherent to NEO-sDFT. Nevertheless, the shape and direction of the polarization can be correctly captured with NEO-sDFT.

\subsection{Large-Scale NEO-FDE Computations} \label{sec:large_scale}

\begin{figure}
	\centering
	\includegraphics[width=0.9\textwidth]{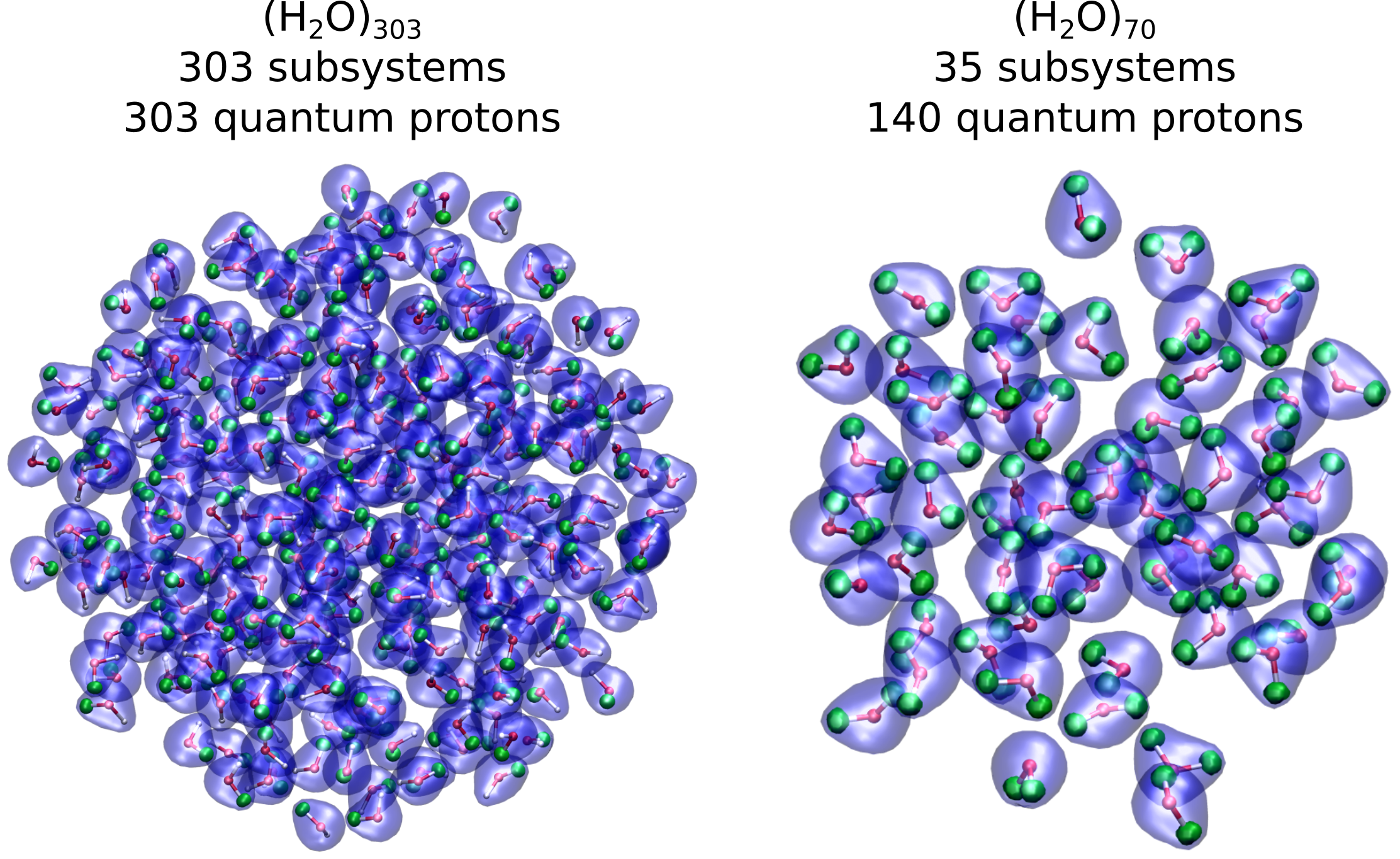}
	\caption{NEO-FDE electron (in blue) and proton (in green) densities computed for the water clusters (H$_2$O)$_{303}$ (left) and (H$_2$O)$_{70}$ (right). The cluster (H$_2$O)$_{303}$ was partitioned into 303 subsystems each being composed of a single water molecule and containing one quantum proton. (H$_2$O)$_{70}$ was split into water dimers with all protons being treated quantum mechanically. The isovalue used is $\pm$0.06~a.u.}
	\label{fig:dens}
\end{figure}

\begin{figure}
	\centering
	\includegraphics[width=0.99\textwidth]{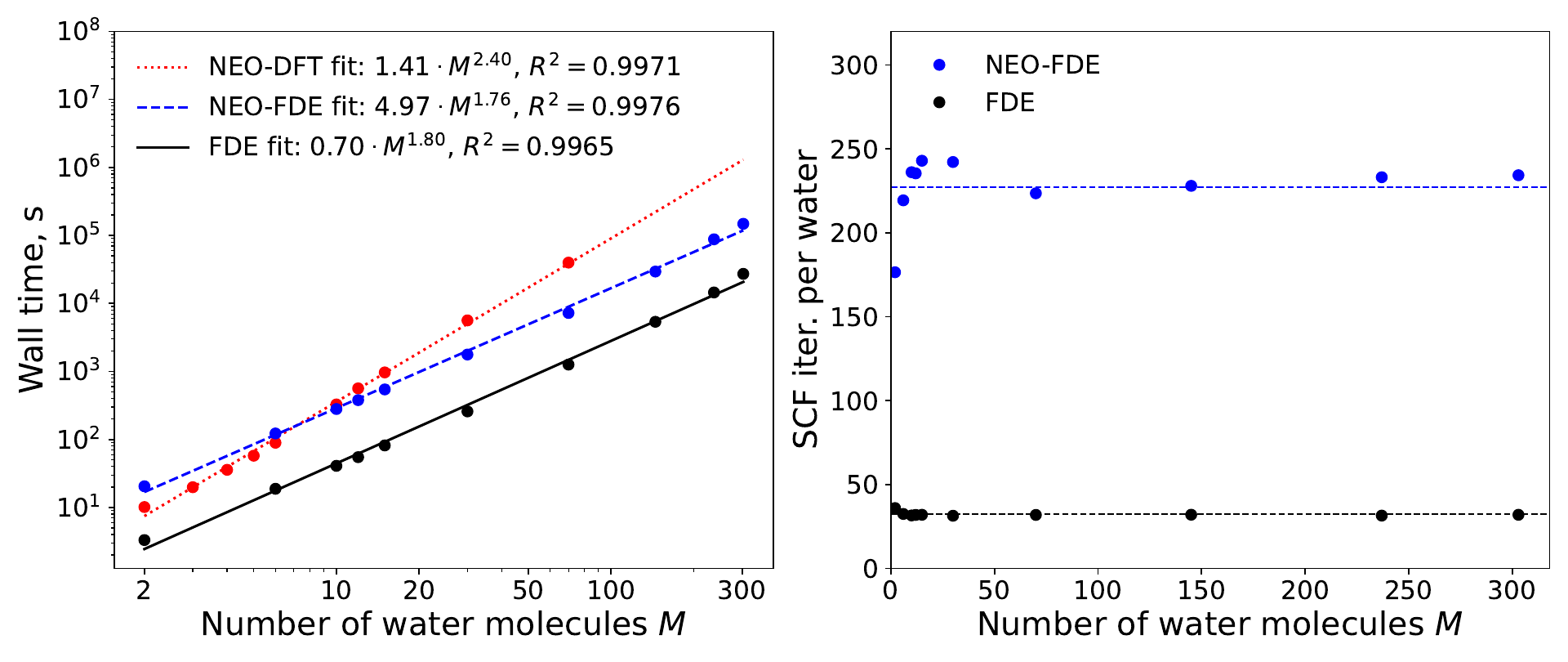}
	\caption{FDE, NEO-FDE, and NEO-DFT wall times on double logarithmic scale (left) and numbers of (NEO-)SCF iterations per water molecule (right). FDE and NEO-FDE results are shown for clusters of 2 to 303 water molecules, and NEO-DFT results for clusters of 2 to 70 molecules. Three freeze-and-thaw cycles were performed in the FDE and NEO-FDE calculations, whereas the NEO-DFT calculations were run for 50 SCF iterations. The wall times were fitted to $C M^A$, where $M$ is the number of water molecules and $C$ and $A$ are fitting coefficients, by linear regression in log--log space. The original data points are shown with circles. Average numbers of SCF iterations are shown with dashed horizontal lines.}
	\label{fig:comp_scaling}
\end{figure}

With the reliability of the NEO-sDFT approach being demonstrated above, we now focus on its computational efficiency for large molecular systems. To that end, we no longer perform freeze-and-thaw cycles until the full convergence of electron and proton densities, but instead fix the number of cycles to a predefined value of three.
We refer to this approach as NEO-FDE. Three freeze-and-thaw cycles are generally not sufficient to converge total energies. However, this is not required for the computational scaling tests. 
This number of cycles was chosen merely to keep the computations feasible. The same number of cycles was employed in several previous studies~\cite{arti2021,arti2020a}, where it was shown to produce qualitatively correct results, whereas the previous scaling tests reported in Ref.~\citenum{unsleber2018} used eight cycles. Similarly, a fixed number of 50 SCF iterations was used in NEO-DFT. As test systems, we employ water clusters (H$_2$O)$_M$ with $M$ ranging from 2 to 303. Examples of NEO-FDE computations of electron and proton densities with different partitioning strategies and assignments of quantum protons are shown in Figure~\ref{fig:dens}. The measured wall times for FDE and NEO-FDE computations of (H$_2$O)$_M$, where each subsystem contains a single water molecule and a single quantum proton, as well as NEO-DFT computations featuring the same assignment of quantum protons are shown in Figure~\ref{fig:comp_scaling}~(left), whereas the numbers of SCF iterations per water molecule required in FDE and NEO-FDE are demonstrated in Figure~\ref{fig:comp_scaling}~(right).

We note that a detailed discussion on the computational scaling of the electronic FDE approach can be found in Ref.~\citenum{unsleber2018}. The authors reported nearly linear scaling for small water clusters with $M<20$. However, for larger numbers of subsystems $M$, the scaling was found to be less favorable due to an increased amount of disk I/O and more involved data structures. The most expensive step was reported to be the evaluation of non-additive functionals, which is performed on the supermolecular grid. Thus, using the pair of XC and non-additive kinetic energy functionals PW91/PW91k for a set of test systems with $M \in \{20, 100, 280\}$, the authors reported the computational scaling of $\mathcal{O}$($M^{1.72}$). In the current implementation of NEO-FDE, no additional non-additive potentials are computed, since the non-additive proton exchange--correlation $\upsilon_{\mathrm{pxc}}^{\mathrm{nad}}$, proton kinetic $\upsilon_{\mathrm{kin}}^{\mathrm{p, nad}}$, and electron--proton correlation $\upsilon_{\mathrm{epc}}^{\mathrm{nad}}  $ potentials are all neglected. The most expensive step, the evaluation of non-additive functionals, is therefore expected to have the same computational cost as in FDE. The additionally evaluated proton--proton $\upsilon_{\mathrm{Coul}}^{\mathrm{pp}}$ and electron--proton $\upsilon_{\mathrm{Coul}}^{\mathrm{ep}}$ Coulomb contributions scale similarly to their electronic counterparts and are expected to increase the prefactor without affecting the exponent. A further increase in the prefactor is expected from the evaluation of electron--proton correlation potentials $\upsilon_{\mathrm{epc}}$ within each subsystem and from the typically larger number of NEO-SCF iterations. We further note that electron and proton densities are evaluated on the same integration grid, so that no additional grids need to be constructed.

As seen from Figure~\ref{fig:comp_scaling}~(left), the fitted exponent of FDE is equal to 1.80 and agrees well with the value of 1.72 reported in Ref.~\citenum{unsleber2018}. Similarly, the exponent of 2.40 obtained for NEO-DFT without density fitting is close to those reported for standard DFT (2.00 with cached integrals and 2.69 without caching)~\citenum{unsleber2018}. As expected, the scaling exponent of NEO-FDE is equal to 1.76 and thus very similar to that of electronic FDE, whereas its prefactor is about 7 times larger. Comparing raw wall times, NEO-FDE is about 5--7 times as expensive as electronic FDE. At the same time, NEO-FDE requires a 5--8 times larger number of SCF iterations. Importantly, as seen from Figure~\ref{fig:comp_scaling}~(right), the number of SCF iterations per water molecule, summed over 3 fixed freeze-and-thaw cycles, does not increase with system size and is on average equal to 32 and 227 for FDE and NEO-FDE, respectively. The higher cost of NEO-FDE can thus be attributed largely to the increase in the number of SCF iterations. Compared to NEO-DFT, for which a fixed number of 50 SCF iterations was used, NEO-FDE becomes faster for clusters containing more than 10 water molecules. Despite its higher cost compared to FDE, NEO-FDE already now enables computations of molecular systems containing up to one thousand atoms, while improved SCF convergence and further improvements of the algorithm, such as implementation of density fitting techniques, are expected to extend its applicability to even larger systems.

It is worth noting that mean-field NEO approaches exhibit slower and more difficult SCF convergence than their conventional electronic-structure  counterparts~\cite{Bochevarov2004,Aodong2022,Feldmann2023,chow2026} making computations of molecular systems containing multiple quantum protons particularly challenging. 
Although computations of each subsystem SCF in NEO-sDFT and NEO-FDE can also pose difficulties, partitioning the molecular system into subsystems limits the number of quantum protons treated in each individual SCF cycle. As a result, hundreds of quantum protons can easily be computed with NEO-sDFT using relatively simple approaches such as the simultaneous DIIS algorithm~\cite{Aodong2022} employed in this work.
To further speed up the individual subsystem SCF, second-order optimization algorithms~\cite{Feldmann2023,Greiner2026} and the removal of nuclear self-Coulomb and self-exchange terms~\cite{chow2026} can be implemented. The latter approach was shown to reduce the number of SCF iterations in NEO-HF by about a factor of five. For NEO-DFT with multiple quantum protons, the combination of this approach with the geometric direct minimization algorithm resulted in a more than twofold reduction in the number of SCF iterations compared to the conventional DIIS-based approach~\cite{chow2026}. Since the computational cost of NEO-sDFT is, as was shown above, largely affected by slow SCF convergence, the implementation of the aforementioned techniques can reduce the scaling prefactor and make computations of larger molecular systems possible, whereas the exponent is expected to remain unchanged. Furthermore, the improved robustness of the nuclear Hartree product representation with respect to the proton initial guesses~\cite{chow2026} can make subsystems containing multiple quantum protons more practical.

Finally, we assess the convergence of NEO-sDFT with respect to the number of freeze-and-thaw cycles for clusters with $M=15$, 30, 70, and 145. The results are shown in Table~\ref{tab:fat_cycles}. While a threshold of $1\times10^{-6}$~a.u.\ was used for all energy computations in Section~\nameref{sec:water_dimers}, looser thresholds of $1\times10^{-5}$~a.u.\ and $5\times10^{-5}$~a.u.\ had to be employed for larger clusters. As one can see from Table~\ref{tab:fat_cycles}, NEO-sDFT can be converged to a density change of 5$\times 10^{-5}$~a.u., requiring a somewhat larger number of 
freeze-and-thaw cycles than sDFT. 
The total energy change between the last two iterations was also found to be below 5$\times 10^{-5}$~a.u.\ in all cases.
The number of required cycles does not increase systematically with system size. Therefore, we expect a similar scaling behavior for fully converged NEO-sDFT calculations as that found for NEO-FDE.

\begin{table}[!h]
	\centering
	\caption{Numbers of freeze-and-thaw cycles required to converge sDFT and NEO-sDFT computations of water clusters (H$_2$O)$_M$ with $M=15$, 30, 70, and 145 for different density convergence thresholds given in a.u.}
	\label{tab:fat_cycles}
	\begin{tabular}{lccc}
		\toprule
		& \multicolumn{2}{c}{sDFT} & NEO-sDFT \\
		\cmidrule(lr){2-3} \cmidrule(lr){4-4}
		$M$ & $5\times10^{-5}$ & $1\times10^{-5}$ & $5\times10^{-5}$ \\
		\midrule
		15  & 6 & 6 & 7  \\
		30  & 6 & 7 & 7  \\
		70  & 6 & 8 & 15 \\
		145 & 6 & 7 & 10 \\
		\bottomrule
	\end{tabular}
\end{table}

\section{Conclusions and Outlook} \label{sec:conclusions}

In this work, we presented the new NEO-sDFT computational approach capable of simulating large molecular systems composed of thousands of atoms while treating all electrons and selected protons quantum mechanically. This was made possible by 
partitioning the total molecular system into subsystems based on the electron and proton densities and employing embedding potentials to account for the inter-molecular interaction. The central feature of our program implementation is its high flexibility allowing one to consider any number of quantum protons per subsystem effectively enabling all types of embedding schemes ($X$)-in-($Y$), where $X$, $Y=$ DFT and NEO-DFT.

The reliability of our approach was assessed by computing interaction energies of a water dimer for different intermolecular displacements and varying numbers of quantum protons, as well as total energies of water clusters containing up to ten molecules.
It was demonstrated that, similar to the conventional electronic sDFT, the non-additive kinetic energy of electrons is among the largest error sources of NEO-sDFT. The deviations between NEO-DFT and NEO-sDFT were found to be about 1--5~meV for the equilibrium separation of the water dimer and up to 63~meV for shorter distances. For larger water clusters, the deviation in total energies grows with system size and reaches about 0.1~eV for ten water molecules, remaining comparable to those of electronic sDFT. Furthermore, NEO-sDFT was shown to capture the polarization of proton densities by the environment, although the polarization is somewhat underestimated compared to NEO-DFT.

Subsequently, we demonstrated the efficiency of our approach by computing water clusters (H$_2$O)$_M$ of increasing size up to $M=303$. Therefore, the largest computation performed treated 303 protons and 3030 electrons quantum mechanically. Additionally, computations with varying sizes of subsystems and quantum proton assignments were shown. The computational scaling with respect to the number of subsystems $M$ was found to be about $\mathcal{O}$($M^{1.76}$). These results agree well with previous studies on the computational scaling of the electronic FDE approach from Ref.~\citenum{unsleber2018}, whereas about seven times larger scaling prefactor can largely be attributed to an increased number of SCF iterations in NEO-FDE compared to electronic FDE. 
We note that our code is not fully optimized and does not yet support density fitting techniques for quantum protons. Therefore, while molecular systems composed of about one thousand atoms are already within reach of NEO-sDFT, further improvements of the computational protocol and the associated scaling are possible.

In conclusion, our results demonstrate that NEO-sDFT provides a robust and accurate framework for accounting for nuclear quantum effects in large molecular systems beyond the size accessible with NEO-DFT. The associated program implementation of our method will soon be available open source. Future extensions of NEO-sDFT towards the construction of simultaneously proton and electron localized states, similar to the FDE-diab methodology~\cite{arti2018,eschenbach2021}, can enable computations of vibronic couplings making an important step towards more realistic simulations of proton transfer and proton-coupled electron transfer processes in complex chemical environments.

\begin{acknowledgement}
	This work was supported by the German Research Foundation (Deutsche Forschungsgemeinschaft, DFG), project number 545861628. Computational resources were provided by the HPC Service of FUB-IT, Freie Universit\"at Berlin. 
\end{acknowledgement}

\newpage
\clearpage
%\bibliography{acs-achemso}

\providecommand{\latin}[1]{#1}
\makeatletter
\providecommand{\doi}
  {\begingroup\let\do\@makeother\dospecials
  \catcode`\{=1 \catcode`\}=2 \doi@aux}
\providecommand{\doi@aux}[1]{\endgroup\texttt{#1}}
\makeatother
\providecommand*\mcitethebibliography{\thebibliography}
\csname @ifundefined\endcsname{endmcitethebibliography}
  {\let\endmcitethebibliography\endthebibliography}{}

\end{document}